\documentclass{article} 
\usepackage{colm2024_conference}

\usepackage{titlesec}
\titlespacing*{\section}{0pt}{0.36\baselineskip}{0.4\baselineskip}
\titlespacing*{\subsection}{0pt}{0.16\baselineskip}{0.2\baselineskip}
\titlespacing*{\subsubsection}{0pt}{0.1\baselineskip}{0.14\baselineskip}

\usepackage{microtype}
\usepackage{hyperref}
\usepackage{url}
\usepackage{booktabs}
\definecolor{darkblue}{rgb}{0, 0, 0.5}
\hypersetup{colorlinks=true, citecolor=darkblue, linkcolor=darkblue, urlcolor=darkblue}
\usepackage{enumitem} 
\usepackage{amsmath}
\usepackage{mdframed}
\usepackage{wrapfig}
\usepackage{tabularx}
\usepackage{listings}
\usepackage{xcolor} 

\title{\vspace{-0.2in}ScenicNL: Generating Probabilistic Scenario Programs\\from Natural Language}

\author{
Karim Elmaaroufi\textsuperscript{1}, Devan Shanker\textsuperscript{1}, Ana Cismaru\textsuperscript{1}, Marcell Vazquez-Chanlatte\textsuperscript{2}, \\
\textbf{Alberto Sangiovanni-Vincentelli\textsuperscript{1}, Matei Zaharia\textsuperscript{1}, \& Sanjit A. Seshia\textsuperscript{1}} \\
\textsuperscript{1}University of California, Berkeley \\
\textsuperscript{2}Nissan Advanced Technology Center - Silicon Valley \\
\texttt{elmaaroufi@berkeley.edu}
}

\usepackage{graphicx}

\colmfinalcopy 
\begin{document}

\maketitle


\begin{abstract}

For cyber-physical systems, including robotics and autonomous vehicles, mass deployment has been hindered by fatal errors that occur when operating in rare events. 
To better understand failure modes, companies meticulously recreate rare crash events in simulation, but current methods do not easily allow for exploring "what if" scenarios which could reveal how accidents might have been avoided.
We present ScenicNL, an AI system that generates probabilistic scenario programs from natural language. Given the abundance of documented failures of autonomous vehicles due to regulatory requirements, we apply ScenicNL to police crash reports, providing a data-driven approach to capturing and understanding these failures.
By using a probabilistic language such as Scenic, we can clearly and concisely represent such scenarios of interest and easily ask “what if” questions. 
We demonstrate how commonplace prompting techniques with Large Language Models are incapable of generating code for low-resource languages such as Scenic.
We propose an AI system via the composition of several prompting techniques to extract the reasoning abilities needed to model probability distributions around the uncertainty in the crash events.
Our system then uses Constrained Decoding and tools such as a compiler and simulator to produce scenario programs in this low-resource setting.
We evaluate our system on publicly available autonomous vehicle crash reports in California from the last five years and share insights into how we generate code that is both semantically meaningful and syntactically correct. Finally, we release our \href{https://ke7.github.io/ScenicNL}{code} and a collection of over 500 crash reports from the California Department of Motor Vehicles.

\end{abstract}

\section{Introduction}

Robust verification and validation of the safety and performance of autonomous artificial intelligence (AI) enabled cyber-physical systems (CPS) requires an integrated approach spanning formal modeling, simulation, and scalable verification and testing methods~\citep{seshia-cacm22a}. Simulation-based verification requires performing an extensive and diverse set of simulations including scenarios that are more dangerous and unusual than typical scenarios. 
As an example, for autonomous driving systems, this allows for evaluation of safety in higher-risk scenarios that may not appear as frequently in on-road testing or log-based simulations. 
Within this domain, there exists a plethora of detailed driving data available through datasets such as nuScenes \citep{caesar_nuscenes_2020} or Berkeley Deep Drive 100K \citep{yu_bdd100k_2020}. 
Recent works such as that by \citet{adewopo2024big} integrate some of these existing datasets into a large video dataset from multiple sources such as dashcams, sensors, traffic surveillance, and small clips from selected video streaming platforms. 

Despite the abundance of publicly-available data, only a fraction contains the corresponding sensor data, and an even smaller portion is relevant to the autonomous driving domain (e.g. traffic surveillance video may not help an autonomous vehicle to be a better driver). 
Moreover, the high-risk situations most important to validate for autonomous driving systems often appear the least frequently in real-world recorded data. 
These scenarios include rare events, and unsafe or unusual driving behaviors by surrounding vehicles, which may result in collisions or dangerous scenarios. 
These underrepresented scenarios are important for testing not only to train systems to avoid them, but they are also necessary for ethical research into decision making when a system must act when a collision is unavoidable \citep{Robinson2021EthicalCA}.
While datasets such as Car Crash \citep{car_crash_dataset_2020} and Dashcam Accident \citep{dashcam_dataset} have collected these higher-risk scenarios, they are only videos without sensor data and are limited in size and duration. For example, the Car Crash Dataset has 1500 examples that are 2 seconds long thus missing the critical events leading up to a crash. 
To the best of our knowledge, there is no dataset containing full driving scenarios in which the recorded vehicles were involved in an accident or in a challenging situation. 
Thus, one promising but challenging opportunity is to generate a realistic dataset containing dangerous driving scenarios.

\begin{figure}[h]
    \begin{center}
        \centerline{\includegraphics[width=\columnwidth]{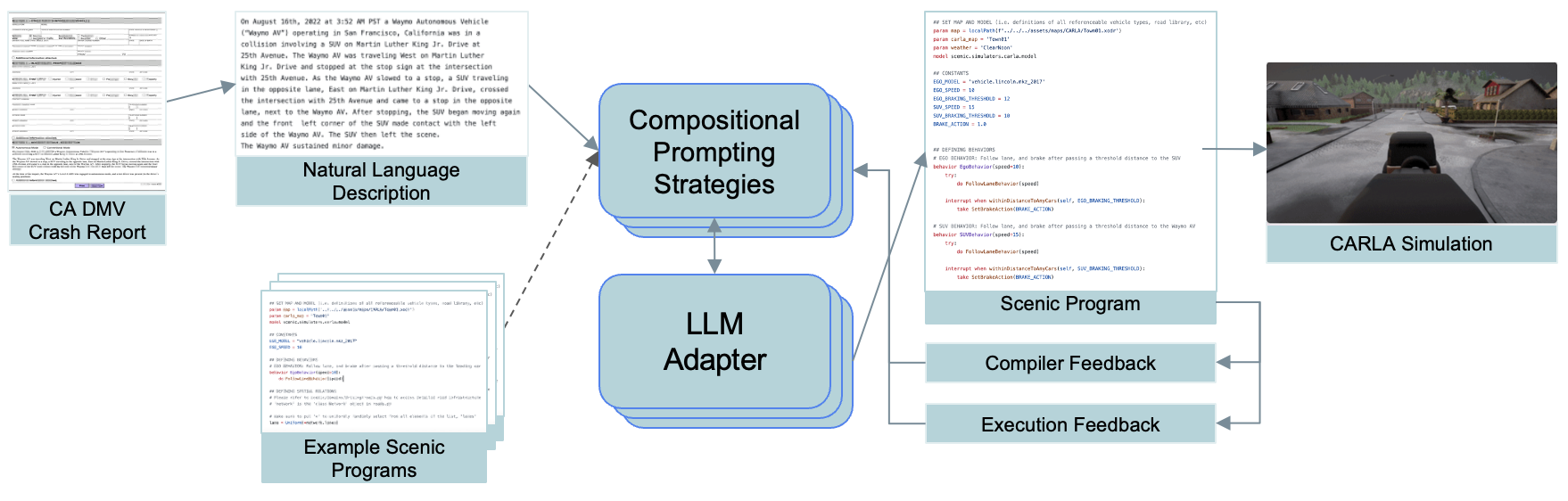}}
        \caption{Architecture diagram of the ScenicNL compound AI system.}
        \label{arch}
    \end{center}
\end{figure}

Given the critical need for simulation based testing and the lack of data pertaining to the rare events needed to gain confidence in our autonomous driving systems, we ask \textit{how can we get this data}?
We observe that within the United States, regulators offer a dataset of a different form than that of sensor or video recordings. For example, the California Department of Motor Vehicles mandates that all AVs must report all crash incidents. The National Transportation Safety Board and the Nevada State Legislature also provide reports for crashes of significant monetary value. Thus we ask, \textit{how can we leverage available reports to reconstruct dangerous and unusual automotive situations in simulation}?
%
%
This question was asked by \citet{scanlon_waymo_2021}. The authors collected all Waymo crashes in Arizona over 10 years and paid a team of "trained experts" to perform exact crash reconstruction. 
Their dataset of 72 crashes were then used to perform counterfactual "what-if" simulations to learn insights about these rare events. 
Similar to Waymo, we reconstruct crash scenes in simulation. 

\begin{figure}[t]
    \vspace{-0.1in}
    \begin{center}
        \centerline{\includegraphics[width=\columnwidth]{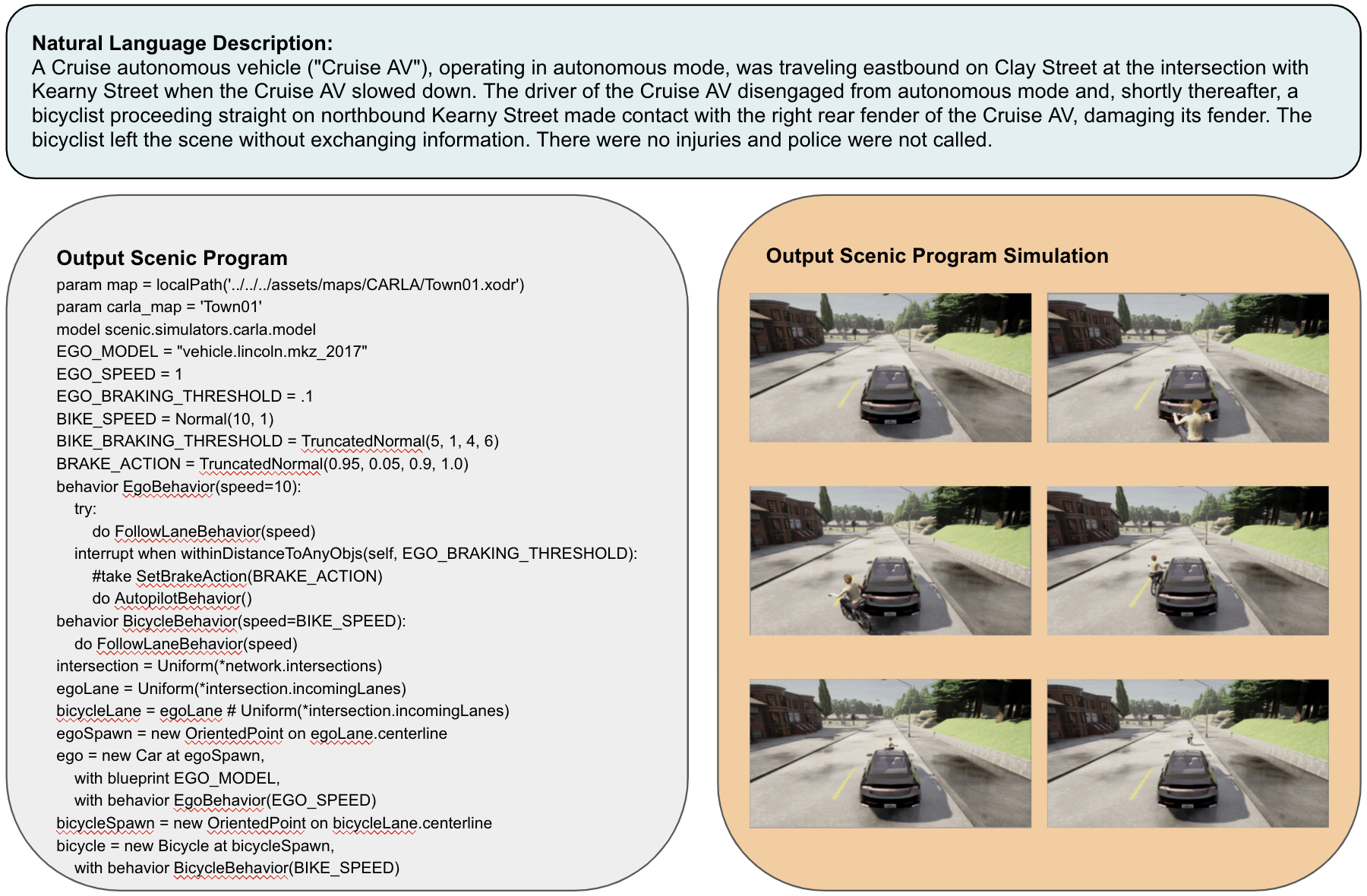}}
        \caption{Example ScenicNL output and its respective CARLA simulator visualization. The natural language description was parsed from a California DMV Crash report. It specifies the collision between a cyclist and paused autonomous vehicle. 
        }
        \label{bicycle_example}
    \end{center}
\end{figure}

In this paper, we address the gap of a missing dataset that contains dangerous and rare driving scenarios by introducing ScenicNL, a system capable of automatically generating such a dataset. 
Our system generates probabilistic programs representing these scenarios of interest from natural language descriptions. 
By generating probabilistic programs, our system can account for uncertainty and variability in the descriptions by modeling events and properties as probability distributions i.e. a single program can represent many variations of the same scenario.
These scenario programs when sampled yield concrete instantiations in a driving simulator.
Unlike prior works, we perform this reconstruction in a completely automated fashion. 
Moreover, we are able to generate programs for a wide variety of scenarios by foregoing coding templates that past works have used. 

In order to produce these scenario programs without templates or human experts, our system leverages Large Language Models (LLMs). 
The challenge in developing ScenicNL is thus, how we can get LLMs to write code in Scenic, a Domain Specific Language (DSL). 
We demonstrate how commonly used prompting techniques are insufficient to write code in our DSL.
Without relying on an extensive knowledge base (vector storage) or fine-tuning, we develop a compound artificial intelligence (AI) system \citep{zaharia_shift_nodate} and share general insights on how the composition of multiple prompting strategies and the integration of tools can help tackle the problem of LLMs writing code for any DSL.
The choice of Scenic as the target language is best illustrated with an example.


\subsection{Scenic as a domain specific language}
Figure~\ref{bicycle_example} illustrates the Scenic program generalizing a collision between a vehicle and a bicyclist.
Scenic itself is a Python-like language augmented with several language constructs to make representing such scenarios easier. For example, Scenic includes several constructs for specifying that an object is visible, on a particular orient point, or relative to another object / point. Additionally, one can declaritively specify distributions -- with all objects modeled by Scenic, e.g. cars, bicycles, weather, etc., coming with prior distributions. For example, the distribution of car models and colors attempts to match known vehicle statistics.
Finally, Scenic enables declaring the behavior of other road users via a co-routine like API.

All together, Scenic enables succinctly capturing common automotive scenarios. This has several benefits. First, the resulting programs are often human readable and editable. Second, because the programs are succinct, they are more likely to fit into the context of a language model. Third, Scenic is an open-source system that is well-supported by verification and testing tools such as VerifAI~\cite{verifai-cav19}. Finally,
because Scenic has a formal semantics, we can guide existing LLMs to generate Scenic programs by using domain specific compiler feedback.

\subsection{\vspace{-0.1in}Contributions}

To scale the reproduction of Scenic programs, ScenicNL uses compositional prompting techniques, a compiler, and Large Language Models (LLM) to develop these programs in a low resource setting such as Scenic. 

Our contributions are: 
\begin{enumerate}[noitemsep,topsep=0pt,leftmargin=0.2in]

    \item We collect all California AV DMV crash reports for the last five years. The reports are classified into easy, medium, and hard. Our system takes these reports and attempts to represent the scenarios as Scenic programs simulated through the CARLA simulator \citep{dosovitskiy2017carla}.

    \item Our system, ScenicNL, is a Compound AI System \citep{zaharia_shift_nodate} leveraging LLMs for generating Scenic Probabilistic Programs from natural language text. Unlike \citet{scanlon_waymo_2021}, our system is fully automated and does not require human experts. It can handle a variety of natural language from simple text descriptions to full PDF crash reports thus surpassing the capabilities of existing diffusion-based approaches such that by \citet{zhong2023languageguided}. Furthermore, our system does is not restricted by template based strategies such as those by \citet{deng_target_2023} and \citet{wang_adept_2023}. It can theoretically represent any scenario. With additional natural language descriptions of AVs, we can represent any scenario and not just those with crashes. 

    \item Qualitative Analysis of our system reveals the difficulty of generating DSLs for LLMs. We share insights on how composing multiple strategies can alleviate some of the difficulties. Our methods are not unique to Scenic, and do not assume the existence of labeled data thus enabling the opportunity for other DSL generation. 

    \item We further augment our approach by taking a systems' approach. We incorporate feedback from existing tools such as the Scenic Compiler to further guide LLMs towards semantically and syntactically correct Scenic programs. Our system consistently outputs syntactically-correct programs at rates of up to $90\%$.

\end{enumerate}

\begin{mdframed}[nobreak=true]
    While our contributions are demonstrated on the autonomous driving domain, we emphasize that Scenic itself is a language applicable to {\em any} AI-enabled CPS or robotics domain, and our ScenicNL system is also independent of the specific application we demonstrate our results on.
\end{mdframed}

\section{Related Work}

There has been much work on reconstructing driving scenarios in simulation for a variety of reasons. Similar to our goals of reconstructing crashes for safety reasons, \citet{kolla_simulation-based_2022} reconstruct crashes in simulation from dash-cam videos. Their methods are incredibly accurate, however, they do require point-cloud data of the incident scene and take three days to reconstruct the scene. In contrast, our system can generate a Scenic program in about 2 minutes. 
\citet{yang_emernerf_2023} forgo traditional simulators and use self-supervision to construct  Neural Radiance Fields (NeRF) for both dynamic objects and static scenes. While visually appealing, NeRFs may not be physically accurate and leave many datapoints absent that otherwise a traditional simulator offers. \citet{kumar2022city} use simulated dashcam views for the purposes of estimating traffic flow (rather than crashes) in CARLA. Works such those by \citet{makridis_adaptive_2023, wang_cadsim_2023}, and \citet{rafati_fard_new_2017} can reconstruct vehicle trajectories using GPS data and priors on vehicle physics, however, they are limited by their datasets which do not contain crash scenarios.

%

Code generation or program synthesis using LLMs has been a popular use case for modern LLMs. Models such as CodeGen \citep{nijkamp_codegen_2023}, CodeLlama \citep{roziere_code_2023}, and StarCoder \citep{li_starcoder_2023} are all examples of recent foundational models that have been pretrained on open source code and usually evaluated on the HumanEval coding dataset \citep{chen_evaluating_2021}. Some models such as CodeGen even have specific fine-tuned versions for Python only code. While these models can write syntactically correct code for popular general purpose languages such as Java or Go, their performance is limited to in distributions tasks just like any machine learning model.

Most of these model's are fine-tuned variants of a foundational model that was trained on a wider body of text such as Common Crawl. For instance Codex is "a GPT language model fine-tuned on publicly available code from GitHub" and it's authors, "study its Python code-writing capabilities" \citep{chen_evaluating_2021}. The authors used 159GB of Python files. Such a vast amount of data is not available for many DSL such as Lean \citep{platzer_lean_2021}, and Isabelle \citep{DiskPaxos-AFP-isabelle}. In these cases, fine-tuning is not an option as the number of examples is ever so small as compared to the amount used in prior works. In the case of Isabelle, there are less than 1000 archived proofs \citep{noauthor_archive_nodate}. 
Thus, new techniques are required. 
One such work is LeanDojo \citep{yang_leandojo_2023}. The authors collect nearly 100k proof tactic examples, and train a small transformer from scratch to output individual proof tactics at every state of a proof. 
While their method works for an interactive theorem prover, it is not able to write a full proof on its own. 
Furthermore, on an even less commonly used DSL, such as Scenic, there are less than 30 publicly available code examples thus making both fine-tuning and the LeanDojo approach (a ReProver model) infeasible.

\section{ScenicNL}
\label{sec:ScenicNL:input}
While works such as~\citet{kolla_simulation-based_2022} have demonstrated that exact crash reconstruction is possible,~\citet{scanlon_waymo_2021} showed that it is when we ask "what-if" questions that we gain insights on how the accident occurred and how we can avoid it. 
Scenic~\citep{fremont_scenic_2019}, a Probabilistic Programming Language (PPL), strikes a balance between the two. With a concise syntax, Scenic allows one to represent the spatial-temporal relationships between objects such as cars on a map. Uncertainty about the environment and object properties can be encoded as probability distributions thus allowing for a natural way to represent uncertainty and "what-if" questions. Scenic also allows for program evaluation with realistic physics by interfacing with simulators. In our setting, we choose CARLA~\citep{dosovitskiy2017carla}, a vehicle simulator illustrated in our introductory example, Figure~\ref{bicycle_example}.
By representing scenarios as Scenic programs and using ScenicNL to generate these programs, we can create hundreds of scenarios answering thousands of "what-if" questions.


\textbf{Problem setup.} 
For simplicity, we assume that the inputs and outputs of our system will belong to a common alphabet $\Sigma$, e.g., tokenized ASCII symbols. We say that a string $P\in \Sigma^*$ is a \emph{Scenic program} if it conforms to the formal grammar/syntax defined in~\cite{fremont_scenic_2019}. We say that a Scenic program $P$ is \emph{valid} if it produces no runtime errors. Let $\text{SCENIC}$ denote the set of all valid Scenic programs. The goal of ScenicNL is to sample valid Scenic programs, $P\in \text{SCENIC}$, given a natural language prompt $x\in\Sigma^*$.

We note two features about this problem: First, because Scenic is a DSL well suited to modeling scenarios for autonomous systems and robotics, the code generation needed is often 
shorter than a general purpose language. On the other hand,
because Scenic is not very common in internet scale datasets,
there is an apriori low probability that an LLM will
generate valid Scenic.

In the next section, we will describe an architecture that incrementally generates a sequences of outputs, $x_1, \ldots, x_n$, by mutating a valid Scenic program $x_i\in \text{SCENIC}$ and checking if it is still valid, $x_{i+1}\in \text{SCENIC}$. 
%
This approach enables the system to understand and bridge the semantic gap between natural language and a Domain-Specific Language (in this case Scenic) in a completely unsupervised manner.

\subsection{System Architecture}
ScenicNL generates Scenic programs from a collection of publicly available crash reports involving AVs. By running a single command, the user can specify any supported LLM as a backbone and compositional prompting strategies. The modular system design supports integration of any new LLM or prompting strategy with the addition of a single new adapter file. The system automatically caches any intermediate outputs to conserve resources and API calls. An architecture diagram is provided in Figure \ref{arch}

\subsection{Input: Autonomous Vehicle Collision Reports}



The system input is a web-scraped collection of over 500 collision reports involving AVs over the last five years, hosted by the California Department of Motor Vehicles \citep{cali-crash-reports}. Scenarios detailed in the autonomous vehicle collision reports may involve interactions with other vehicles, pedestrians, cyclists, or other objects. ScenicNL uses optical character recognition (OCR) to extract any relevant data from the collision report PDFs. The final natural language input contains collision information including weather and environmental conditions, sustained vehicle damage, and description of the events leading up to the accident. The input data represents a vast range of scenarios with potentially multiple cars, pedestrians, several modes of locomotion (e.g. skateboarding), and collisions.
Next, due to the wide variety of scenarios possible, we manually classify the reports:

\begin{enumerate}[noitemsep,topsep=0pt,leftmargin=0.2in]

    \item \textbf{Easy} Reports that contain one dynamic agent (e.g. an AV was rear-ended by another agent while the AV was legally at rest) on a public road.
    \item \textbf{Medium} There are two dynamic agents (e.g. an AV and cyclist), or the scenario is in a road configuration not supported in a default CARLA map (e.g. a parking lot).
    \item \textbf{Hard} There are three or more dynamic agents, scenarios involving aggressive drivers, or scenarios involving advanced driving maneuvers (e.g. illegal vehicle overtaking).

\end{enumerate}


\subsection{Prompting}

\label{sec:prompting}
\begin{figure}[h]
    \begin{center}
        \centerline{\includegraphics[width=\columnwidth]{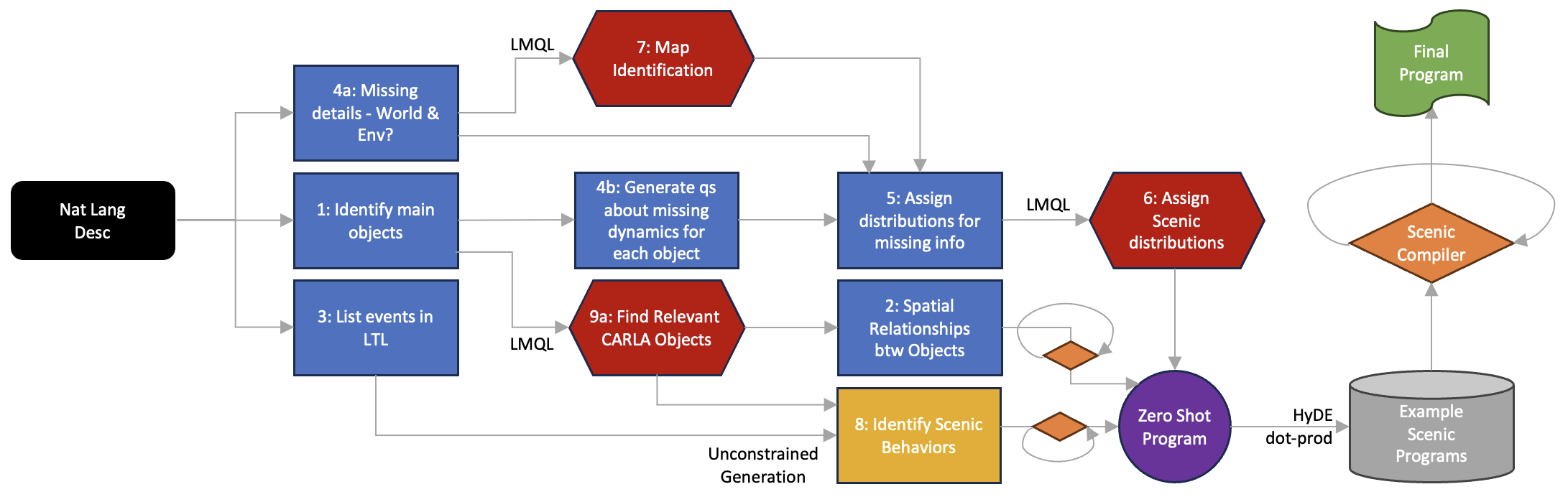}}
        \caption{Overview of ScenicNL's Compositional Prompting. Blue rectangles utilize Tree-of-Thought with Role Playing and a single example program. Red hexagrams use Constrained Decoding to generate correct Scenic code. A single-shot program is generated from partially correct code, reasoning in natural language, and program subparts which were corrected (if necessary) with a compiler in the loop. A Few-Shot full program is then generated and looped with a compiler to emit a final program.
        }
        \label{comp_prompts}
    \end{center}
    \vskip -0.2in
\end{figure}

\textbf{Modular LLM Prompting}
ScenicNL accepts the natural language scene description as input in order to formulate a prompt given a user-specified approach and language model. The modular LLM prompting mechanism allows for ease of evaluation and code generation on a wide range of approaches. 
This may range from unique decoding constraints to few-shot examples retrieved from a vector database. 
ScenicNL allows users to specify a model and prompting technique in the one-line \texttt{gen\_scenic} command. 
The full list of strategies supported are described in detail in the Appendix.

\textbf{Compositional LLM Prompting}
Some prompting techniques are better for different reasons at different times. For example, few-shot prompting can be used to demonstrate a specific output format whereas, chain-of-thought \citep{wei2023chainofthought} can be used to elicit better reasoning capabilities from an LLM. Our key insight is that we should separate reasoning from coding as much as possible. When working in the natural language space, we can employ multiple personas to further elicit world knowledge from a single LLM thus better identifying and describing uncertainties in a scenario. Furthermore, we ask these personas to generate their own questions about uncertainly in the crash report descriptions which they then literately attempt to answer. By using multiple personas and self-generated questions, we find that the quality and quantity of identified uncertainties from a report increases over a single LLM query. We also observe that leveraging constrained decoding to help guide the LLMs to syntactically correct code where possible further improves system performance as we don't have to hope for correct code which must be validated by a compiler downstream. We implement the following composition of prompts:

\begin{enumerate}[noitemsep,topsep=0pt,leftmargin=0.2in]

    \item \textbf{Tree of Thought (ToT)} \citep{yao2023tree} We ask the LLM to play the role of 3 crash reconstruction experts. Each one must identify the relevant dynamic and static objects in the scene. They debate amongst each other which objects are relevant and provide a final agreed upon answer.  

    \item \textbf{Zero-Shot} Given the list of relevant objects and the original description, we ask the LLM to ask questions about the missing properties of those objects (speed, placement, and type). We also ask questions about the environment (weather and what kind of road setting does the scenario take place in).

    \item \textbf{ToT} We again return to our experts to debate about the answers of the prior questions. Each question is answered directly where possible (e.g. weather condition or road structure such as 4 way intersection) or in the form of a probability distribution, where confidence may be represented as variance (e.g. Vehicle speed can be represented as a Gaussian distribution with mean and variance).

    \item \textbf{Few-Shot \& Constrained Decoding} Given that our experts were able to reason about the relevant objects and uncertainty in the original description, we now construct the Scenic program by parts (constants/parameters, agent behaviors, and spatiotemporal relations). These program parts are further fed into the next parts of the ScenicNL system. The constrained decoding is implemented through LMQL \citep{lmql} — a package that allows users to express constraints as python code rather than having to directly manipulate token logit biases. We use LMQL version 0.7 with \texttt{gpt-3.5-turbo-instruct}, the most recent OpenAI model compatible with LMQL custom constraints. Finally, we leverage Retrevial Augmented Generation (RAG) \& Hypothetical Document Embeddings (HyDE) in each section to further help the LLM. We refer the reader to the appendix for a deeper explanation on RAG and HyDE.
    \vspace{-0.15in}
\end{enumerate}


An overview of our prompts is shown in Figure \ref{comp_prompts}. The full prompts and questions from the figure are provided in the appendix.

\subsection{\vspace{-0.05in}Scenic Tools}
Each of the program sub parts are fed into the Scenic compiler to check for syntactic correctness. 
If a subpart fails, a zero-shot prompt with the compiler error message and prior context is passed to the LLM to correct the error up to a certain number of retries.
Note that because Scenic is built on top of python, there is no type checking or reference checking during compilation.
Once all parts pass, they are stitched together into a final program. 
This final program is then executed up to a certain number of tries where again a zero-shot prompt is used to explain any execution errors to the LLM.
An interesting future direction would be to instead use the abstract syntax tree of the program since one can represent a syntactically correct program through a JSON schema thus paving way for a variety of Constrained Decoding tools apart from LMQL, including direct input into OpenAI's API.

\subsection{\vspace{-0.05in}Output: CARLA Simulation Execution}
For all successful end-to-end runs, ScenicNL can generate multiple simulations by sampling a single fixed program from the Scenic program. Each sample yields concrete values which are drawn from the distributions defined in the programs. This is then run in a simulator (in our case CARLA, however, Scenic does support other simulators). Each simulation is then rendered with realistic physics, dynamics, and graphics. This allows for further semantic validation of the ScenicNL output. Since our output is a concise program, asking "what-if" questions is as simple as modifying a small portion of the program. 

Our initial experiments with GPT-4 Vision and LLaVA, indicate that a strong Vision-Language Model (VLM) such as GPT-4 can serve as an adequate vision critic. However, we do not include the vision critic in our final system as we have yet to determine where in the system the feedback should be provided (i.e. directly to the Scenic program or prior in the natural language space during the ToT reasoning). This is left as future work.

With the programs that are generated, there are numerous downstream applications including using tools such as VerifAI \citep{verifai-cav19} to perform formal analysis of their learning based components to better understand their system's failure modalities, learning Operational Design Domains \cite{opdesigndo, torfah_odd} to predict when an accident is imminent, retraining of Reinforcement Learning policies on the new data generated by these programs, and many others.

\section{\vspace{-0.1in}Evaluation}

\textbf{Baseline Evaluations}
ScenicNL was evaluated for all supported configurations of language models and prompting techniques. The results are indicated in Table \ref{results-table}. In addition to evaluating widely used prompting techniques, we also selected DSPy \cite{khattab_dspy_2023} as an example of an automatic prompt-optimization system. We utilized the \texttt{BootstrapFewShot} and \texttt{BootstrapFinetune} \texttt{teleprompts} with \texttt{answer\_exact\_match} as a metric or loss function. As it's highly unlikely a full program can be exactly matched, we split the program into five parts (parameter defintions, scene setup, behaviors, placement, and post conditions) and apply the loss function to each part. This loss function, even by parts, leaves much to be desired and an interesting future direction would be to create a new one that can account for both semantic and syntactic correctness. With Constrained Decoding, we use LMQL and write constraints as Python functions. With Function Calling, we define our own API as an intermediary way to build a Scenic program. You can find further details about each prompting technique in the Appendix \ref{prompt_strats}.

The evaluation process incorporated the following qualitative and quantitative metrics:

\begin{itemize}[noitemsep,topsep=0pt,leftmargin=0.2in]

    \item \textbf{Syntactic Correctness:} assesses if the generated code is valid Scenic code that uses proper syntax and that compiles.
    
    \item \textbf{Accuracy:} the extent to which the output program is semantically similar to the natural language input
    
    \item \textbf{Relevance:} the amount of unnecessary input content that the model was able to successfully filter out. 
    
    \item \textbf{Expressiveness:} the extent to which the output code leverages the full capabilities of Scenic and includes complex code constructs and probabilistic features.

    \item \textbf{ARE Score:} the average of the Accuracy, Relevance, and Expressiveness metrics. A measure of the extent to which the output program appropriately models the input scenario, leveraging the tools available in Scenic.
    
    \item \textbf{Inference Time:} assesses the inference time to generate one output measured in seconds per program.
    
\end{itemize}

Note that the qualitative metrics of Accuracy, Relevance, and Expressiveness rely on human experts familiar with Scenic to evaluation the program outputs. They are measured out of five and the rubric is provided in the Appendix \ref{rubric}. Since all three qualitative factors are imperative in generating useful Scenic code, we propose the ARE score as a method of fairly weighting all of these factors when evaluating our Scenic programs.

\begin{table*}[ht]
\centering
\caption{Baseline evaluation of simple prompting techniques. No compiler or execution feedback was used to improve outputs further. Inference time is measured in seconds per output. For techniques in which more than one model was evaluated, we average their results as there was no statistical significance between them. With DSPy, the BootstrapFewShot and BootstrapFinetune methods were used with exact string match as the loss function.
\\\scriptsize{° Evaluated with Anthropic claude-instant-1.2. * Evaluated with OpenAI gpt-3.5-turbo-0613. \textsuperscript{†} Evaluated with OpenAI gpt-3.5-turbo-instruct.}}
\label{results-table}
\begin{tabularx}{\textwidth}{lXXXXXX}
\toprule
& Correct-ness & Accuracy & Relevance & Expressive-ness & ARE Score & Inference Time \\
\midrule
Zero-Shot°*\textsuperscript{†} & 1 & 2.3 & 3.8 & 2.2 & 2.8 $\pm$ 0.1 & 1.064 \\
Few-Shot°*\textsuperscript{†} & 11 & 3.1 & 2.5 & 2.7 & 2.8 $\pm$ 0.2 & 0.718 \\
Function Call.* & 0 & 2.4 & 2.1 & 1.5 & 2.0 $\pm$ 0.2 & 5.411 \\
Con. Decoding\textsuperscript{†} & 26 & 3.2 & 3.5 & 3.9 & 3.5 $\pm$ 0.4 & 8.023 \\
RAG° & 1 & 3.9 & 2.9 & 3.6 & 3.4 $\pm$ 0.2 & 0.766 \\
RAG + HyDE°\textsuperscript{†} & 7 & 3.8 & 3.4 & 3.5 & 3.6 $\pm$ 0.1 & 1.537 \\
DSPy\textsuperscript{†} - Exact match & 0 & - & - & - & - & - \\
ScenicNL & 90 & 4.1 & 4.5 & 4.2 & 4.3 $\pm$ 0.1 & 109.860 \\
\bottomrule
\end{tabularx}
\end{table*}

\subsection{\vspace{-0.1in}Compositional Prompting Evaluation}

Given the classification of input descriptions in \ref{sec:ScenicNL:input} we perform partial human evaluations only on the \texttt{easy} dataset. While parts of the \texttt{medium} are possible, most are not due to the lack of compatible maps in CARLA. In the \texttt{hard} dataset, we have observed that additional prompting is needed to breakdown advanced driving maneuvers (e.g. an overtake can be described as accelerate, switch lane to the left, pass the other vehicle, switch lane to the right, and decelerate). This is required as Scenic does not natively support such maneuvers and instead only supports primitive maneuvers. An alternative future direction could be to extend the behaviors of Scenic to support such high-level driving behaviors.

We begin by walking through our prompting strategy described in Section \ref{sec:prompting} with the example in Figure \ref{bicycle_example}. For brevity, we show the example output of just the dynamic object query, \textit{What are the relevant dynamic objects in the description?}
\begin{verbatim}
EXPERT 1:
1. Cruise autonomous vehicle ("Cruise AV")
2. Bicyclist
3. Right rear fender of the Cruise AV
EXPERT 2:
1. Cruise AV
2. Bicyclist
3. Intersection of Clay and Kearny Streets
EXPERT 3:
1. Cruise AV
2. Bicyclist
Panel Discussion:
All three experts identified the main objects in the 
scene as the Cruise AV and the Bicyclist.
FINAL ANSWER:
1. Cruise AV
2. Bicyclist
\end{verbatim}
Given the relevant objects, we then continue with object by object prompts to find the closest matching objects supported in the chosen simulator (CARLA). Observe that by using ToT, we have limited two possible objects which could have been identified as relevant: the rear fender of the Cruise AV which is in fact part of the Cruise AV, and the intersection which is not an object but rather an environmental property which will be later answered in a different question. 
We repeat the ToT prompting for static objects and environment details. 

One of the main features of Scenic is the ability to quantify uncertainty through distributions. Thus, we ask our experts about things that are unknown from the description. Following each unknown quantity or property, we further ask for possible distributions over those values. An example final answer of unknown properties is:
\begin{verbatim}
The details missing from the description include weather, time of day, traffic 
conditions, road surface conditions, layout of the intersection, presence of 
road markings or signs, and the behavior of the Cruise AV and the bicyclist leading 
up to the accident. These details are important in understanding the environment 
and the sequence of events that led to the accident.
\end{verbatim}
Examining the output of just the bicyclist's behavior, we have: 
\begin{verbatim}
BIKE_SPEED = Normal(10, 1)
BIKE_BRAKING_THRESHOLD = TruncatedNormal(5, 1, 4, 6)
BRAKE_ACTION = TruncatedNormal(0.95, 0.05, 0.9, 1.0)
\end{verbatim}

Still, it is possible that despite providing the immediately necessary information to write Scenic, the LLM may still write incorrect code. 
To that extent, this is where having a compiler check partially completed programs is beneficial. 
Here is one such example where once the entire constants section is populated, the compiler has identified an error:
\begin{verbatim}
...
EGO_MODEL = 'vehicle.lincoln.mkz_2017'
EGO_SPEED = 10
EGO_TURN_SIGNAL = 'left'
TRUCK_MODEL = vehicle.carlamotors.carlacola
TRUCK_SPEED = 10
TRUCK_TURN_SIGNAL = 'straight'

C. The first compiler error raised with the scenic program:
Error details below..
error message: name 'vehicle' is not defined
D. Please output a modified version of scenic_program modified so the 
compiler error does not appear
...    
\end{verbatim}
The LLM correctly identified which built-in CARLA vehicle to use, however, it improperly formatted the truck model as it did not surround the name with quotes. By provided this error to the LLM along with a few examples of typical errors, it is quickly fixed. 

The same techniques of expert discussion and iterating through answers are applied to the agent behaviors, and spatiotemporal relations program sections. However, during the last inference step which generates the code, we additionally augment the prompt with 3 examples which are collected via RAG \& HyDE. This is necessary as primitive Scenic behaviors are often composed to meet a description and spatiotemporal relations knowledge of the properties of the underlying objects. One spatiotemporal example is locating a 4-way intersection on a map and placing a vehicle there:
\begin{verbatim}
fourWayIntersection = filter(lambda i: i.is4Way, network.intersections)
intersec = Uniform(*fourWayIntersection)
startLane = Uniform(*intersec.incomingLanes)
ego_spwPt = startLane.centerline[-1]
ego = new Car following roadDirection from ego_spwPt for DISTANCE_TO_INTERSECTION1,
        with behavior EgoBehavior(trajectory = ego_trajectory)
\end{verbatim}
This is done because the number of properties available are larger than context window of several LLM models. 
While recent advancements in LLMs have resulted in Claude's 200k \citep{claude3fam}, and Gemini's 1 million context window \citep{reid2024gemini}, the "needle in a haystack" identified by \citet{noauthor_llmtest_needleinahaystackreadmemd_nodate} is still a concern if we were to instead provide all possible properties in the prompt.
Thus, one future direction could be to reference the Backus-Naur Form (BNF) grammar defined in \citet{fremont_scenic_2019} and create a corresponding LMQL constraint or GGML BNF grammar as used in \citet{llamacpp}.

\section{Conclusion and Future Work\vspace{-0.05in}}

We propose ScenicNL, a Compound AI System that generates Scenic programs from California DMV crash reports. 
We choose to represent these scenarios as Scenic programs because of the ability to model uncertainty as probability distributions in a probabilistic programming language such as Scenic. 
Additionally, Scenic includes primitives to represent a wide variety of Cyber Physical Systems (CPS) and their environments. 
We demonstrate that simple prompting techniques are not enough for state-of-the-art LLMs to produce correct code in low-resource settings like Scenic.
We build a system that leverages several prompting strategies, and existing tools, to effectively generate programs in this low-resource setting. 
Our techniques are not unique to Scenic and can be used for other DSLs. 

Our work offers several promising future directions. ScenicNL supports the generation and rendering of realistic CARLA simulations from traffic accident reports, which will enable more efficient human-in-the-loop or automated end-to-end validation of autonomous driving systems. 
Exploring how to leverage these simulations in the training of reinforcement-learning based agents is one such direction. 
Other directions include updating LMQL or replacing it with a more expressive and performant constrained decoding library such as AICI \cite{Moskal2024}, and updating Scenic so that we can support more behaviors. Finally, as Scenic supports other CPS, we can extend this work to support other domains like Robotics and Aviation.

\section*{Acknowledgments\vspace{-0.05in}}

This work was supported in part by Provably Correct Design of Adaptive Hybrid Neuro-Symbolic Cyber Physical Systems, Defense Advanced Research Projects Agency award number FA8750-23-C-0080; by Nissan and Toyota under the iCyPhy Center; by C3DTI, and by Berkeley Deep Drive.







\bibliography{colm2024_conference}
\bibliographystyle{colm2024_conference}

\appendix
\section{Appendix}

\subsection{Prompting Strategies}
\label{prompt_strats}
For each approach introduced below, a brief description of the end-to-end process of generating a Scenic program for that approach is provided.

\subsubsection{Zero-shot}
The zero-shot approach involves a brief description of the Scenic programming language adapted from the Scenic documentation. Though the documentation includes several method names and descriptions, the approach does not provide any well-formed sample Scenic programs as part of the input.

\subsubsection{Few-shot}
The baseline few-shot approach involves the same brief description of the Scenic programming language adapted from the documentation. In addition to providing the same Scenic tutorial, the prompting provides several hand-picked examples of Scenic programs written for the CARLA simulator based on the CARLA Autonomous Driving Challenge. The structure and number of few-shot examples provided is consistent across all few-shot prompting techniques in order to preserve consistency across all reported statistics.

\subsubsection{Retrieval Augmented Generation (RAG)}
RAG is a technique that enhances the generation process by adding information from relevant documents within a specialized database as additional context. This augmented few-shot approach adapts the same structure as the baseline few-shot approach. However, instead of using the same few-shot example Scenic programs for every input, the program selects the most semantically similar examples from a vector database containing CARLA Challenge Scenic programs. To evaluate semantic similarity, embeddings (calculated using the all-mpnet-base-v2 model \citep{song2020mpnet, allmpnet}) of the example Scenic programs are stored in a vector database (we use Pinecone) and are compared against individual input prompt embeddings using vector dot product as the distance metric.

\subsubsection{Hypothetical Document Embeddings (HyDE)}
Hypothetical Document Embeddings (HyDE) \citep{gao_precise_2022} intend to improve upon the performance of traditional RAG systems, by converting the model input into the same distribution as the set of data contained in the vector database. This is achieved through an initial generation of a Scenic output using the few-shot technique, followed by performing RAG on the embedding of the baseline few-shot Scenic output instead of the direct natural language input.

\subsubsection{Function Calling}
Prior work in other DSLs demonstrates the improved performance of having a language model write DSL code using a custom code-generation API. A function calling approach API is written in a general-purpose programming language contained within the training distribution of the language model, generating code in the desired DSL output language. More specifically, the OpenAI documentation and other sources \citep{function-calling} recommend a function-calling approach that requests the language model to generate output in a language contained within the domain.

The function calling approach requires the development of an API written in a general-purpose programming language (GPL) that generates lines of valid Scenic code. The LLM is instructed to generate code in the GPL using the API rather than directly generating Scenic. Each line of language model output is called by a GPL interpreter to generate Scenic code. In case any API calls are malformed, as a fallback, the LLM attempts to act as an interpreter for any GPL to Scenic API calls that fail to compile.

\subsubsection{Constrained Decoding}
Constrained decoding involves the specification of constraints that impose significant positive or negative weights on token generation. As a result, we can more strictly ensure specific tokens appear or do not appear, reducing the risks of hallucination or syntax errors. Implementing such restrictions within a programming language is challenging, as variable and function definitions can take any form, potentially bypassing the specified constraints. Our method circumvents this issue by splitting up program generation into different sections, each with a set of structural and token-by-token constraints. This ensures compliance with constraints while maintaining the flexibility needed for creating more diverse programs. Specifically, we implement this using LMQL \cite{lmql} as it allows us to describe constraints through Python code as opposed to having to directly manipulate per token biases. Still, we cannot overly constrain the outputs within each section without falling into a templating based strategy such as that by \citet{wang_adept_2023} or \citet{deng_target_2023}.

\subsubsection{DSPy}
The DSPy framework converts the problem of prompt engineering into a learning problem, to improve the efficiency and performance of the overall process. By framing the task as a learning problem, DSPy is able to optimize the best natural language prompt for a given task. Via this process, DSPy can implicitly leverage many of the techniques described above and several others. In order to do so, DSPy requires a measurable loss function. The ideal loss function would be able to measure the semantic meaning between a program and a natural language description, the program's syntactic correctness, and our proposed ARE metrics. Designing such a loss function would be an interesting future direction. Presently, we use the default \texttt{answer\_exact\_match} which checks for an exact string match. As it's unlikely a correct program can be proposed via an exact string match, we instead breakdown the program into parts and apply the loss function accordingly. Thus, we define the following DSPy Signature:

\begin{lstlisting}[caption={DSPy Scenic Program Signature}]
class GenScenic(dspy.Signature):
    """
    Write Scenic probablistic programs given a natural language description 
    of an autonomous vehicle crash report.
    """
    question = dspy.InputField()
    param_definitions = dspy.OutputField(desc='Necessary imports and parameter definitions')
    scene_setup = dspy.OutputField(desc='Objects that should be placed in the scene')
    behaviors = dspy.OutputField(desc='Dynamics of the objects in the scene')
    placement = dspy.OutputField(desc='Placement of the dynamic objects in the scene')
    post_conditions = dspy.OutputField(desc='Requires and Terminates clauses')
\end{lstlisting}

\subsection{Prompt Details}

As mentioned in Figure \ref{comp_prompts}, we utilize the composition of Tree-of-Thought \cite{yao2023tree} with Role Playing, Zero-Shot Prompting, Few-Shot Prompting, and Constrained Decoding. For the blue rectangles which utilize Tree-of-Thought with a single example, the following meta-prompt is used:

\begin{verbatim}
Scenic is a probabilistic programming language for modeling the environments of 
autonomous cars. A Scenic program defines a distribution over scenes, 
configurations of physical objects and agents. Scenic can also define 
(probabilistic) policies for dynamic agents, allowing modeling scenarios where 
agents take actions over time in response to the state of the world. We use CARLA 
to render the scenes and simulate the agents.
     
We are going to play a game. For the following questions, imagine that you are 
3 different autonomous driving experts. For every question, each expert must 
provide a step-by-step explanation for how they came up with their answer. 
After all the experts have answered the question, you will need to provide a final 
answer using the best parts of each expert's explanation. Use the following format:
"EXPERT_1:"
"<expert_1_answer>"
"EXPERT_2:"
"<expert_2_answer>"
"EXPERT_3:"
"<expert_3_answer>"
"FINAL_ANSWER:"
"<final_answer>

Here is one example of a Scenic program:"
"{example}\n"
     
QUESTION <insert_qs_num>:
\end{verbatim}

The numbers of the questions in Figure \ref{comp_prompts} correspond to how one could think of how they would manually reason about a crash. The exact questions asked in our prompts are as follows (with few-shot examples omitted for brevity, if specified):

\begin{enumerate}
    \item This question utilizes an additional single shot example to demonstrate how to fill out the reasoning template from the meta-prompt. The scenario and text of the single shot example were synthetically generated and human edited. \begin{verbatim}
Based on the description, what are the main objects that need to be 
included in the scene? Provide step-by-step reasoning then provide 
your final answer as a numbered list. Be concise in your reasoning (no 
more than 1-2 sentences per object).
    \end{verbatim}

    \item In this question, we use no examples. We also enforce a token limit. \begin{verbatim}
Based on the relevant objects selected from the original description, what 
are the spatial relationships between the objects? (e.g. car is in front of 
pedestrian, etc.) Are the objects moving or stationary? Are they visible or 
occluded? You can only use the following terms to describe spatial 
relationships: in front of, behind, left of, right of, facing, ahead of, 
behind, visible, and not visible.
Each expert and the final answer should be provided in the following format:
SPATIAL_RELATIONSHIPS:
<spatial_relationships>
MOVEMENT:
<movement>
VISIBILITY:
<visibility>
JUSTIFICATION:
<Q2_JUSTIFICATION>
FINAL ANSWER:
<Q2_FINAL_ANSWER>
    \end{verbatim}

    \item A token limit is enforced. \begin{verbatim}
What are the main events that happened in the scene? (e.g. car stopped when 
pedestrian crossed the street, a car was driving in a lane then switched 
lanes then made a left turn, etc.). Describe these events using Linear 
Temporal Logic (LTL).
    \end{verbatim}

    \item \begin{enumerate}
        \item A token limit is enforced. \begin{verbatim}
What details about the world and environment are missing from the 
description? (e.g. weather, time of day, etc.)
    \end{verbatim}

        \item A 2-shot synthetically generated and human edited example is used with a token limit. \begin{verbatim}
For each of the relevant objects, what details about the objects are 
missing from the description that you would need to ask the author about 
in order to create a more accurate scene? What are the main 
environmental factors that need to be included in the scene? Your 
questions should cover dynamics of objects in motion (e.g. speed), 
distances between every pair of objects, and environmental conditions 
(e.g. weather). Provide your questions as a numbered list, but do not 
ask about personal details of any individuals involved.
        \end{verbatim}
    \end{enumerate}

    \item A single shot synthetically generated and human edited example is used with a token limit. \begin{verbatim}
Based on the missing object information from the user, provide a reasonable 
probability distribution over the missing values. Answer only the questions 
that are about distance between objects, speed, weather, or time. For 
example, if the time of day is missing but you know that the scene is in the 
morning, you could use a normal distribution with mean 8am and standard 
deviation 1 hour (Normal(8, 1)). If the color of the car is missing, you 
could use a uniform distribution over common car color string names. If the 
car speed is missing, you could use a normal distribution with mean around a 
reasonable speed limit for area of the scene and reasonable standard 
deviation, etc.
    \end{verbatim}

    \item A token limit and constrained decoding for the correct objects are enforced. Note that these limits are not actually due to Scenic, but rather CARLA. \begin{verbatim}
A user will provide you with probability distributions for missing 
information in a vehicle crash description. Your task is to interpret the 
probability distributions and express them as a Scenic program.

Scenic can only support the following distributions so you must pick the 
closest matching distribution. Under no circumstance should you use any of 
the other distributions

"Range(low, high) - Uniform distribution over the real range [[low, high]]"
"DiscreteRange(low, high) - Uniform distribution over the discreet integer 
range [[low, high]]"
"Normal(mean, std) - Normal distribution with mean and standard deviation"
"TruncatedNormal(mean, stdDev, low, high) - Normal distribution with mean 
and standard deviation truncated to the range [[low, high]]"
"Uniform(value, …) - Uniform distribution over the list of values 
provided."
"Discrete([[value: weight, … ]]) - Discrete distribution over the list 
of values provided with the given weights (e.g., [[value: 0.5, value: 0.2, 
value: 0.3]])"

"For weather, Scenic can only support a Uniform or Discrete distribution over 
the following values: 
['ClearNoon', 'CloudyNoon', 'WetNoon', 'WetCloudyNoon', 'SoftRainNoon', 
'MidRainyNoon', 'HardRainNoon', 'ClearSunset', 'CloudySunset', 'WetSunset', 
'WetCloudySunset', 'SoftRainSunset', 'MidRainSunset', 'HardRainSunset', 
'ClearNight', 'CloudyNight', 'WetNight', 'WetCloudyNight', 
'SoftRainNight', 'MidRainyNight', 'HardRainNight' ,'DustStorm']"

Based on the distributions and original description, define Scenic 
distributions over the uncertain values. Provide values for the parameters 
to your distributions. You may not use any of the other distributions. If 
you cannot find a distribution that matches the missing information, you 
must choose the closest matching distribution.
    \end{verbatim}

    \item Constrained decoding for the correct town selection and a token limit are used. For the description portion, we do not use constrained decoding. For brevity, we do not repeat the list of default maps and descriptions of CARLA here as they can be found in our code or in CARLA's public documentation. \begin{verbatim}
Based on the original description, pick from the following the best 
matching town. You may not choose any other town. If you cannot find a town 
that matches the original description, you must choose the closest 
matching town. Then after selecting a town, provide a high-level 
description (ignoring road names) of where in the town we should replicate 
the original description. For example, if the original description 
specified a highway such as US-101, provide a description about the 
properties of that highway, such as it is a 4 lane road.
<INSERT_LIST_OF_CARLA_MAPS_AND_DESCRIPTIONS>
    \end{verbatim}

    \item A token limit is enforced. For brevity, we omit the full list of supported Scenic behaviors here as they can be found in our code or in Scenic's public documentation. \begin{verbatim}
Here is a list of the supported behaviors in Scenic. Based on the relevant 
objects and important events, which behaviors do we need to use to recreate 
the original description? You may select more than one behavior as they are 
composable. If you cannot find a behavior that matches the original 
description, you must choose the closest matching behavior.
<INSERT_LIST_OF_SCENIC_BEHAVIORS>
    \end{verbatim}

    \item A single synthetically generated example is provided. A token limit is used. For brevity, we omit the full list of vehicles and objects that CARLA support as they can be found in our code or in CARLA’s public documentation\begin{verbatim}
A user will provide you with a list of main objects from a description. For 
each of the main objects, find the closest matching models from the list 
below. If there are any objects in the original description that you see a 
match for (e.g. a traffic cone), include them in your answer even if they 
are not listed as a main object. Specify your answer as the string value of 
that model. In your final answer, only respond with python code as plain text 
without code block syntax around it.
    \end{verbatim}
\end{enumerate}

\subsection{ARE Score Rubric}
\label{rubric}

We further elaborate on the human evaluation criteria for the Accuracy, Relevance, and Expressiveness scores presented in Table \ref{results-table}. They are measured on a linear scale from 1 to 5 where one is the lowest and 5 is the highest.

\subsubsection{Accuracy}
The extent to which the program output contains semantic content present in the natural language input.

\begin{enumerate}
  \item The program code is too vague or generic to represent useful content from the description, could easily apply to many other scenes.
  \item Several behavior names or assignments hint at scene comprehension, but many remain generic or incorrect.
  \item Some major behavior definitions, assignments, or vehicle classes align with text, but some fail to do so.
  \item Only minor details from natural language scene missing from output Scenic program.
  \item Scenic program captures full extent of essential agents, behaviors, initial conditions involved in natural language input.
\end{enumerate}

\subsubsection{Relevance}
The extent to which the program output omits irrelevant or absent semantic content not present in input.

\begin{enumerate}
  \item The program emphasizes events or street names not relevant to key traffic scenario.
  \item The program references city names, street names, or other constructs or behavior names not accessible in Scenic.
  \item The program contains limited semantic representation of information not present in the natural language in a manner consistent with Scenic.
  \item The program contains only minor artifacts of few-shot prompting or irrelevant behaviors/objects occurring in only smaller program components.
  \item The program demonstrates no indication of semantic influence not contained in the input natural language.
\end{enumerate}

\subsubsection{Expressiveness}
The extent to which the program leverages the full expressive powers of Scenic as a language.

\begin{enumerate}
  \item The program only uses basic behavior definitions and constructs as defined in the docs, with very limited use of Scenic's expressivity.
  \item Behaviors or positional assignments demonstrate hints of understanding basic building blocks of Scenic language.
  \item Generated program leverages main Scenic constructs to represent scene, but neglects multiple sources of more enhanced expressivity.
  \item Program leverages most capabilities of Scenic, but is neglecting a source of key expressive power (distributional objects or behavior definitions or positional assignments).
  \item Program leverages full capabilities of Scenic as language, including enforcing conditions on scene and sampling from distributions.
\end{enumerate}

\end{document}